\pdfoutput=1

\documentclass[11pt,twoside,a4paper,cmspaper,final,collab]{cms-tdr}
\def\svnVersion{249998}\def\cmsCernNo{2014-005}\def\cmsCernDate{\today}\def\cmsMessage{Published in the Journal of High Energy Physics as \href{http://dx.doi.org/10.1007/JHEP06(2014)120}{\doi{10.1007/JHEP06(2014)120.}}}

\begin{document}\cmsNoteHeader{SMP-13-004}

\hyphenation{had-ron-i-za-tion}
\hyphenation{cal-or-i-me-ter}
\hyphenation{de-vices}

\RCS$Revision: 243681 $
\RCS$HeadURL: svn+ssh://svn.cern.ch/reps/tdr2/papers/SMP-13-004/trunk/SMP-13-004.tex $
\RCS$Id: SMP-13-004.tex 243681 2014-05-26 14:58:53Z tdupree $
\cmsNoteHeader{SMP-13-004} 
\title{Measurement of the production cross sections for a Z boson and one or more b jets in $\Pp\Pp$ collisions at $\sqrt{s} = 7\TeV$}

\author{The CMS Collaboration}

\date{\today}

\abstract{
  The production of a Z boson, decaying into two leptons and produced
  in association with one or more b jets, is studied using proton-proton collisions
  delivered by the LHC at a centre-of-mass energy of 7\TeV.
  The data were recorded in 2011 with the CMS detector and correspond to an integrated luminosity of 5\fbinv.
  The Z$(\ell\ell)$+b-jets cross sections (where $\ell\ell$ = $\mu\mu$ or ee) are measured
  separately for a Z boson produced with exactly one jet and with at least two b jets.
  In addition, a cross section ratio is extracted for a \cPZ\ boson produced with at least one \cPqb\ jet, relative to
  a \cPZ\ boson produced with at least one jet.
  The measured cross sections are compared to various theoretical predictions,
  and the data favour the predictions in the five-flavour scheme, where b quarks are assumed massless.
  The kinematic properties of the reconstructed particles are compared with the predictions from the \MADGRAPH event generator
  using the \PYTHIA parton shower simulation.
}

\hypersetup{%
pdfauthor={CMS Collaboration},%
pdftitle={Measurement of the production cross sections for a Z boson and one or more b jets in pp collisions at sqrt(s) = 7 TeV},%
pdfsubject={CMS},%
pdfkeywords={CMS, physics, Standard Model, electroweak, heavy flavor, b jets, b tagging}
}

\maketitle 

\section{Introduction}

Z bosons and jets originating from bottom quarks (b jets) are produced copiously in proton-proton collisions at the Large Hadron Collider (LHC).
The production of a Z boson with at least one b jet in the detector acceptance, Z+b-jets production,
is useful for precision tests of perturbative QCD~\cite{Campbell:2003dd,Maltoni:2005wd,Campbell:2005zv}.
The production of a Z boson with a single b jet, Z+1b-jet production,
provides information relating to the b-quark content of the proton.
The study of the production of a Z boson in association with at least two b jets, Z+2b-jets production,
is of interest since it is a background in many searches for yet unobserved processes,
such as the production of heavier supersymmetry-like Higgs bosons via vector boson fusion,
and in studies of the standard model Higgs boson produced in association with a Z boson and decaying to b quarks~\cite{Hbb,Hbb2}.

The production of a Z boson with b jets
originates in proton-proton collisions
from gluon-gluon and quark-antiquark interactions, the former being the dominant contribution~\cite{Campbell:2005zv}.
A smaller contribution, expected to be less than 5\% based on measurements of the effective area for hard double-parton interactions~\cite{MPI,MPI2}, 
originates from multiple parton interactions (MPIs).
The production cross section for a Z boson with at least one b jet has
been measured previously at the LHC at $\sqrt{s} = 7\TeV$ by the ATLAS~\cite{AtlasZb} and CMS~\cite{Zbpaper} Collaborations
and by the CDF~\cite{CDF} and D0~\cite{D0} Collaborations at the Tevatron $\Pp\Pap$ collider, at $\sqrt{s} = 1.96\TeV$,
where the dominant contribution comes from quark-antiquark interactions.
The characteristics of the production of a Z boson in association with b hadrons
have been studied at the LHC by the CMS Collaboration~\cite{ewk11015}.

In this paper, measurements are reported of the cross sections at $\sqrt{s} = 7\TeV$
for the production of a Z boson with exactly one jet and separately for the production of a Z boson with at least two b jets.
Two event categories are defined according to the b-jet multiplicity, and the yields are corrected for
the respective backgrounds and efficiencies, taking into account possible migrations of events between the two categories.
The cross sections are estimated at the level of stable final-state particles
and are compared with predictions from \MADGRAPH~\cite{MADGRAPH5}
in the five-flavour (5F) scheme, where b quarks are assumed massless, and the four-flavour (4F) scheme, where massive b quarks are used,
as well as with the next-to-leading-order (NLO) predictions from a{\MCATNLO}~\cite{Frederix:2011qg}.
The inclusive Z+b-jets cross section is compared to the production of a Z boson in association with jets of any type.
The resulting ratio has smaller theoretical and experimental uncertainties than the absolute cross section~\cite{D02}
and is used to elucidate the apparent difference between the measured Z+b-jets cross section~\cite{Zbpaper}
and the prediction at the parton level from the {\MCFM} NLO generator~\cite{Maltoni:2005wd}.

In addition, the distributions of reconstructed kinematic observables for jets and leptons in the Z+2b-jets final state are
compared to a Monte Carlo (MC) simulation using the matrix element
calculations of \MADGRAPH in the five-flavour
scheme and using \PYTHIA~\cite{PYTHIA}
for the simulation of the parton shower and hadronization processes.
Understanding the details of the kinematics is important in the search for
undiscovered particles as well as for the study of the newly discovered Higgs boson~\cite{AtlasHiggs,Higgs1,Higgs2} in similar topologies.

\section{CMS detector and event samples}
\label{sec:sel}

The data used in this analysis were collected with the Compact Muon Solenoid (CMS) detector.
The central feature of the CMS apparatus is a superconducting solenoid of $6\unit{m}$ internal diameter that provides a magnetic field of $3.8\unit{T}$.
Within the field volume are a silicon pixel and strip tracker, a crystal electromagnetic calorimeter (ECAL), and a brass/scintillator hadron calorimeter.
Muons are detected in gas-ionisation detectors embedded in the steel flux return yoke of the magnet.
A more detailed description of the CMS detector can be found elsewhere~\cite{CMS:2008zzk}.
A right-handed coordinate system is used in CMS, with the origin at the nominal interaction point,
the $x$ axis pointing to the centre of the LHC ring and the $y$ axis pointing up, perpendicular to the plane of the LHC ring.
The polar angle $\theta$ is measured from the positive $z$ axis, which points along the anticlockwise beam direction,
and the azimuthal angle $\phi$ is measured in the $x$-$y$ plane.
The pseudorapidity is given by $\eta=-\ln[\tan(\theta/2)]$.

The data were collected in 2011 at a proton-proton centre-of-mass energy of 7\TeV
and correspond to an integrated luminosity of $\mathcal{L} = 5.05 \pm 0.11\fbinv$~\cite{CMSLumi}.
During the course of data taking, the instantaneous luminosity
increased from $10^{32}$ to $3.5\times10^{33}\percms$,
resulting in an average number of proton-proton interactions per bunch crossing (pileup) of 9.7 with an RMS of 4.7.

Events are selected using dimuon and dielectron triggers.
The dimuon trigger \pt thresholds were increased from 7\GeV on both muons to 13 and 8\GeV on the leading and subleading muons, respectively,
as the instantaneous luminosity increased during the data taking period~\cite{MuReco}.
The dielectron trigger has transverse momentum (\pt) thresholds of 17 and 8\GeV, loose identification criteria,
and very loose isolation requirements~\cite{EReco}.

In order to compare the data to the theoretical expectations,
signal events and the expected backgrounds (Z+jets, $\ttbar$, and ZZ) are
generated by MC simulation and
simulated within the CMS detector using \GEANTfour~\cite{geant}.
Inclusive Z+jets and $\ttbar$ events are simulated with \MADGRAPH 5.1.1.0,
using \PYTHIA 6.424 with the Z2 tune~\cite{Z2,Z2_2} for the parton showers, hadronization, and MPIs.
The CTEQ6L1 parton distribution functions (PDFs)~\cite{cteq} are used.
The ZZ sample is simulated using \PYTHIA.
The Z+jets sample is also used to extract the signal efficiencies
and for the comparison of kinematic distributions.

The simulated samples used for comparison with data are normalized to the cross sections expected from theory in the full acceptance.
The cross section for the Z+jets sample, 3048\unit{pb}, is normalized to match the next-to-NLO prediction for inclusive Z production obtained
with \textsc{fewz}~\cite{DY_NNLO} and the CTEQ6m PDFs~\cite{cteq}.
NLO predictions obtained from \textsc{mcfm} are used for the normalization of the {\ttbar} sample, 157.5\unit{pb},
and the ZZ sample, 6.2\unit{pb}~\cite{ZZMCFM}.
The simulated Z+jets sample is split into three subsamples, according to the
underlying production of b jets, c jets, or jets originating only from
gluons or u,d,s quarks (hereafter called light-parton jets), with no
requirement on the \pt or $\eta$ of the jets. These subsamples are
labelled Z+b, Z+c, and Z+l, respectively.

\section{Event reconstruction and selection}

The reconstruction and selection of events with a Z boson that decays into a pair of muons or electrons, and one or more b jets
are based on the criteria used in the
measurement of the inclusive Z+b-jets cross section at CMS~\cite{Zbpaper}.
For the identification of muons, jets, and missing transverse energy, the CMS particle-flow event
reconstruction is used.
This algorithm combines the information from all subdetectors to identify
and reconstruct the individual particles produced in the collision~\cite{CMS_PAS_PFT-09-001,CMS_PAS_PFT-10-001}.

The leptons in the analysis are required to originate from the primary vertex,
which is chosen as the vertex with the
largest quadratic sum of the \pt of its constituent tracks.
Muons are reconstructed by combining the information from both the silicon tracker and the muon spectrometer in a global fit.
Tight requirements, including particle-flow identification, are applied to the muon candidates to ensure high purity~\cite{MuReco}.
Electrons are identified by combining tracker tracks and ECAL clusters, including the
ECAL deposits from bremsstrahlung~\cite{EReco}.
An isolation variable, which is defined as the sum of the
magnitudes of the transverse momenta of the particles reconstructed
in a cone around the lepton candidate, $\DR=\sqrt{\smash[b]{(\Delta\eta)^2+(\Delta\phi)^2}}<0.4$ (0.3),
relative to the transverse momentum of the lepton,
is used to reject muons (electrons) that are embedded in jets.
Charged particles not associated with the primary vertex
are not considered in forming the isolation variable.
To reduce the effect from pileup,
the contribution of neutral particles is corrected by subtracting the energy deposited in the
isolation cone by charged particles not associated with the primary
vertex, multiplied by a factor of 0.5.
This factor corresponds approximately
to the ratio of neutral to charged hadron production
in the hadronization process of pileup interactions~\cite{MuReco,EReco}.
After this correction, the isolation variable is required to be less than 20\% for muons and 15\% for electrons.

Both leptons are required to have
$\pt^\ell > 20\GeV$ and pseudorapidity $|\eta^\ell| < 2.4$.
Opposite charges for the leptons are required when forming pairs.
In the case of multiple lepton combinations, the lepton pair with the
invariant mass closest to the nominal Z-boson mass is selected as the Z candidate.
The efficiency of the dilepton selection is estimated
using the tag-and-probe method~\cite{CMS_paper_WZ} in
events with at least two leptons and a jet passing the requirements
detailed below.
The offline selection efficiencies are estimated from data and simulations,
and data/simulation `scale factors' are estimated to correct for the differences;
trigger efficiencies are estimated from data alone.
All simulated events are corrected for differences between data and simulation by
applying the trigger efficiencies and the data/simulation scale factors as a function of \pt and $\eta$ for each lepton.

Jets are reconstructed by clustering individual particle-flow objects
using the anti-\kt jet clustering algorithm~\cite{Salam} with a distance parameter of
0.5, as implemented in the \textsc{fastjet} program~\cite{Cacciari:2005hq,FastJet}.
Jets are calibrated using photon+jet, Z+jet,
and dijet events to ensure a uniform energy response in \pt and $\eta$~\cite{CMS_PAS_JME-10-001}.
The contribution to the jet transverse
energy from pileup is estimated on an event-by-event basis using the
jet-area method~\cite{cacciari-2008-659} and is subtracted.
The reconstructed jets are required to have $\pt^j > 25\GeV$ and to be
separated from each of the selected leptons by at least $\DR(\ell,j)=0.5$.
Furthermore, jets are required to have $\abs{\eta^j} < 2.1$ to ensure
optimal b-tagging performance. Loose identification
criteria~\cite{CMS_PAS_JME-10-001} are applied in order to
reject jets coming from beam background, calorimeter noise, and isolated photons.
Jets originating from pileup in the Z+jets sample, and thereby contributing falsely to the cross section ratio,
are suppressed by requiring the momentum of particle tracks originating from the selected primary vertex
compared to the jet momentum be at least 10\%.
The remaining background caused by jets from pileup is ${\sim}2\%$ in the Z+jets data sample.

Jets originating from b quarks are tagged by taking advantage of the
long b-hadron lifetime. The `Simple Secondary Vertex' (SSV) b-tagging algorithm
employs a three-dimensional flight
distance significance between the primary vertex and a secondary vertex in a jet.
To maximise the selection efficiency of the Z+b-jets process for multiple b jets,
the high-efficiency version of the SSV b-tagging algorithm is used,
which considers secondary vertices built from two or more tracks.
The discriminant value to define b-tagged jets is chosen such that the probability of
tagging a light-parton jet (mistagging fraction) is less than 1\%, with a b-tag efficiency of ${\sim}55\%$.
The b-tagging efficiencies and mistagging fractions are measured in the
data and simulation as functions of the \pt and $\eta$ of the jet using
inclusive jet samples,
where the tagging efficiency in the data is ${\sim}5\%$ smaller
than the efficiency in the simulations~\cite{CMS_PAS_BTV-11-002}.
Simulated events are corrected for this difference,
taking into account the data/simulation scale
factor for each b-tagged jet, depending on the generator-level flavour.

\begin{figure}
  \begin{minipage}{0.48\linewidth}
    \includegraphics[width=1.0\linewidth]{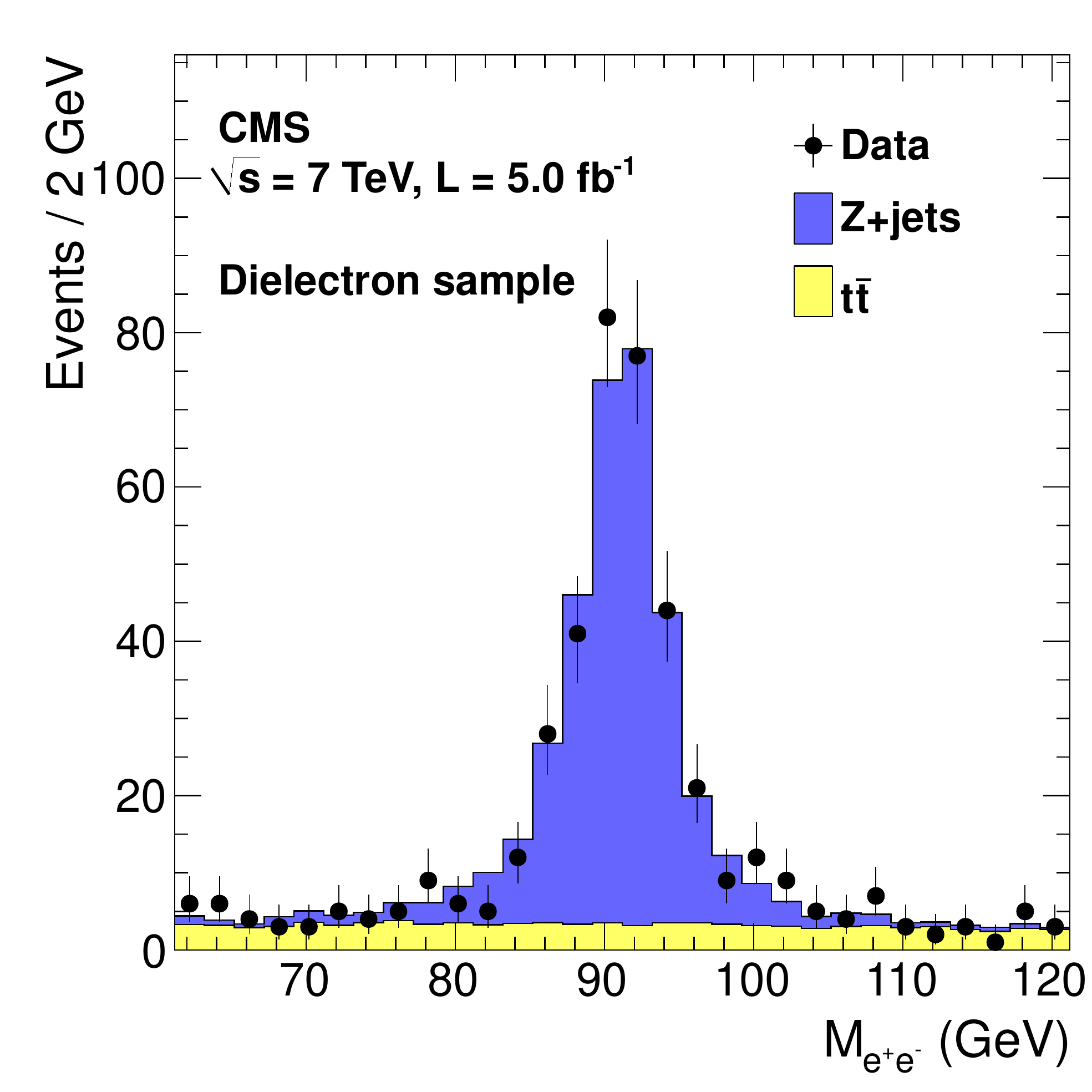}
    \caption{
	Distribution of the invariant mass of the electron pair in a sample of events containing
        two electrons and two b-tagged jets and requiring $\MET\text{ significance}<10$.
        Overlaid are the distributions after a fit of the \ttbar fraction within the wide dilepton invariant-mass window: $61<M_{\ell\ell}<121\GeV$.
    }
    \label{fig:mll}
  \end{minipage}
  \hfill
  \begin{minipage}{0.48\linewidth}
    \includegraphics[width=1.0\textwidth]{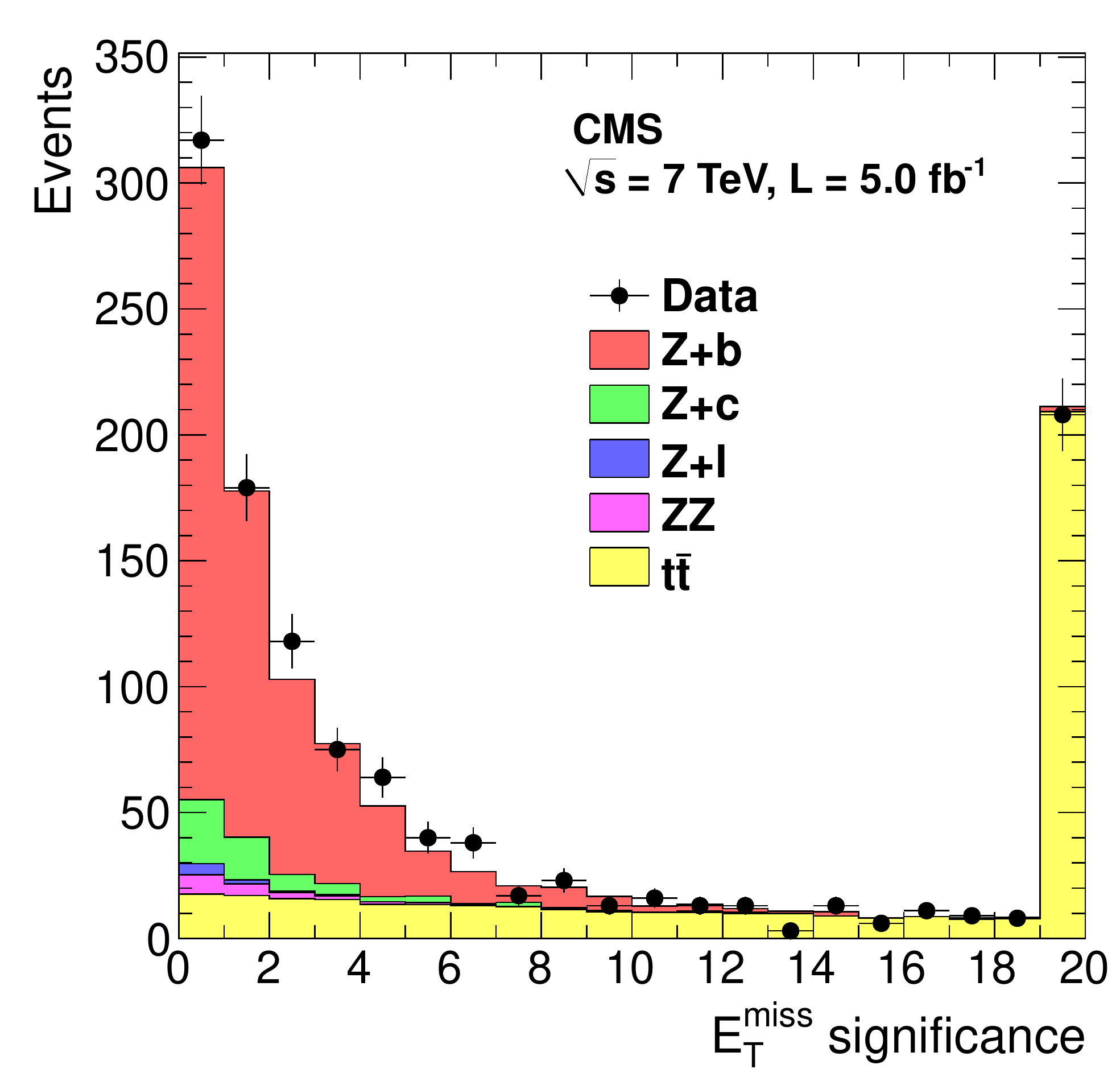}
    \caption{
	Distribution of the \MET significance variable in a sample of events containing two leptons and two b-tagged jets
        and within the default mass window, $76<M_{\ell\ell}<106\GeV$.
        The simulated distributions are normalized using the theoretical predictions. The last bin contains the overflow.
}
    \label{fig:met}
  \end{minipage}
\end{figure}

After the application of the b-tagging requirement,
the sample is divided into nonoverlapping categories according to the number of b-tagged jets in the sample:
the Z+1b-jet sample contains events with exactly one b-tagged jet, while the Z+2b-jets sample contains the events with at least two b-tagged jets.
In order to suppress background from \ttbar production in both samples,
the reconstructed dilepton invariant mass $M_{\ell\ell}$ is required to have a value
between 76 and 106\GeV.
In Fig.~\ref{fig:mll} the dielectron invariant mass distribution shows the effectiveness of this requirement.

To further suppress the \ttbar background in the Z+2b-jets sample,
the missing transverse energy (\MET) is evaluated and events with a value significantly different from zero are vetoed.
The \MET is calculated by forming the negative vector sum of
the transverse momenta of all particles in the events.
The \MET significance is more robust than the \MET itself against pileup,
and offers an event-by-event assessment of the likelihood that the observed \MET is consistent with zero
given the reconstructed content of the event and known measurement resolutions of the CMS detector~\cite{CMS_PAS_JME-10-009}.
In Fig.~\ref{fig:met} the \MET significance distribution is shown after requiring a
Z candidate and two b-tagged jets.
The distributions for the Z+b and \ttbar components motivate the selection of events with a reconstructed \MET significance less than 10,
which results in a high signal efficiency and small systematic uncertainty.

All simulated events are corrected for the differences between data and simulation
in the pileup distributions, b-tagging efficiencies, and lepton reconstruction efficiencies.
The data yields as well as the predicted yields are summarised in Table~\ref{tab:yields}.

\begin{table}[t!h]
\begin{center}
\topcaption{{
Data yields in the selected samples
and a comparison to the expectation from various sources based on MC simulations.
The expected yields are estimated using the theoretical predictions for the cross sections. Uncertainties are statistical only.
}}
\label{tab:yields}
{\footnotesize
\begin{tabular}{l| r r|r r r r r }
\hline
Selection              &    Data   & Total simulation  & Z+b             & Z+c             & Z+l             & \ttbar        & ZZ               \\
\hline
Z($\mu\mu$)+1b-jet     & $13\,090$ & $12\,904\pm77$    & $6810\pm58$     & $3647\pm41$     & $1829\pm29$     & $549\pm3$     & $69\pm1$     \\
Z($\mu\mu$)+2b-jets    &   $522$   &   $480\pm13$      &  $350\pm12$     &   $34\pm4$      &    $5\pm1$      &  $80\pm1$     & $11\pm1$  \\
\hline
Z(ee)+1b-jet           &  $9672$   &  $9924\pm67$      & $5218\pm50$     & $2844\pm36$     & $1364\pm25$     & $445\pm3$     & $53\pm1$  \\
Z(ee)+2b-jets          &   $362$   &   $357\pm11$      &  $258\pm10$     &   $27\pm3$      &    $2\pm1$      &  $62\pm1$     &  $8\pm1$    \\
\hline
\end{tabular}
}
\end{center}
\end{table}

\section{Backgrounds}
\label{sec:bkg}

Events not originating from the Z+b-jets production process, but nevertheless
contributing to the final reconstructed event yield after the full
selection, are expected to originate from \ttbar, Z+jets, and ZZ
production.
For the Z+1b-jet sample, the main background originates from Z bosons produced in association with non-b jets;
for the Z+2b-jets sample, another sizable background originates from \ttbar production, with another nonnegligible contribution from ZZ production.

The background originating from \ttbar production is estimated
by means of a binned fit to the wide dilepton invariant-mass spectrum,
$61<M_{\ell\ell}<121\GeV$,
as shown for the electron channel in Fig.~\ref{fig:mll}.
The shape of the invariant-mass distribution for Z+jets events
is taken from Z-boson-enriched data samples,
while the distribution (template) for \ttbar is based on simulation.

As a control method, two other distinct parameterizations are employed for the probability density functions of the Z+jets and \ttbar contributions:
(i)~Z+jets templates based on simulation together with \ttbar templates based on distributions in data samples, and (ii)~an empirical parameterization.
These \ttbar templates are acquired from an opposite-flavour ($\mu/\Pe$) dilepton sample in which the \ttbar contribution is enriched.
The empirical parameterizations employ a relativistic Breit--Wigner distribution to describe the Z+jets contribution
and a polynomial distribution to describe the \ttbar contribution;
the parameters of both probability density functions are free to vary in the fit.
As another control method, a multivariate matrix-element approach~\cite{MADWEIGHT} is used
to distinguish signal and background.

In all channels the results obtained with the various parameterizations and methods are consistent with each other and with the expectations from simulation.
The fraction of events from \ttbar, $f_{\ttbar}$,
estimated from the fit within the wide mass window
is interpolated to the signal mass window ($76<M_{\ell\ell}<106\GeV$).
The differences between the \ttbar estimates derived from alternative parameterizations are used
to estimate the related systematic uncertainty.

The background due to mistagged c and light-parton jets is estimated from the
mass distribution of the secondary vertices ($M_\mathrm{SV}$) of the b-tagged jets.
For the Z+1b-jets sample, exactly one jet per event is b-tagged, and hence one secondary vertex per event is reconstructed and analyzed.
For the Z+2b-jets sample the distributions of the $M_\mathrm{SV}$ of both the leading (in \pt)
and subleading b-tagged jets are used.

\begin{figure}[t!]
  \begin{center}
    \includegraphics[width=0.48\textwidth]{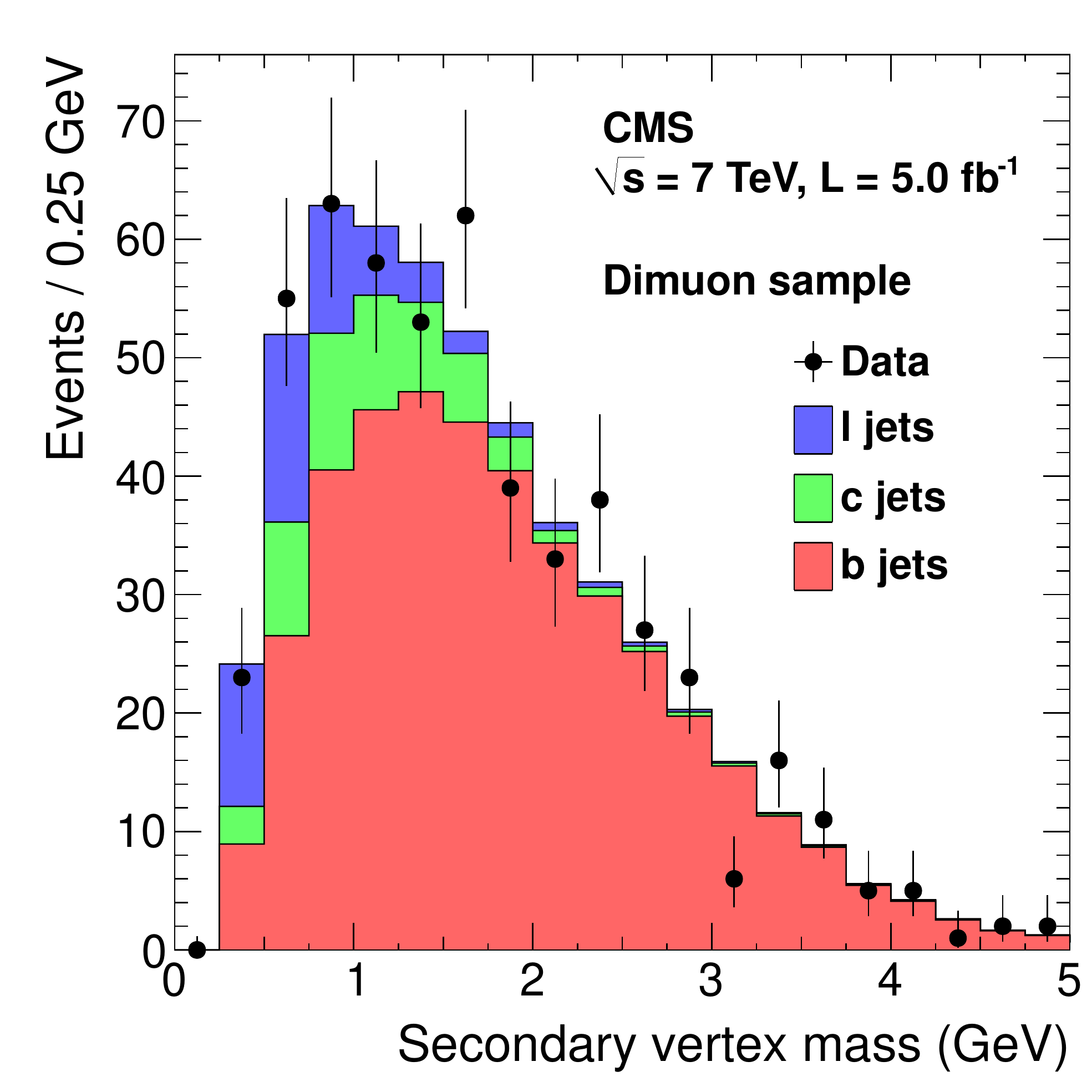}
    \includegraphics[width=0.48\textwidth]{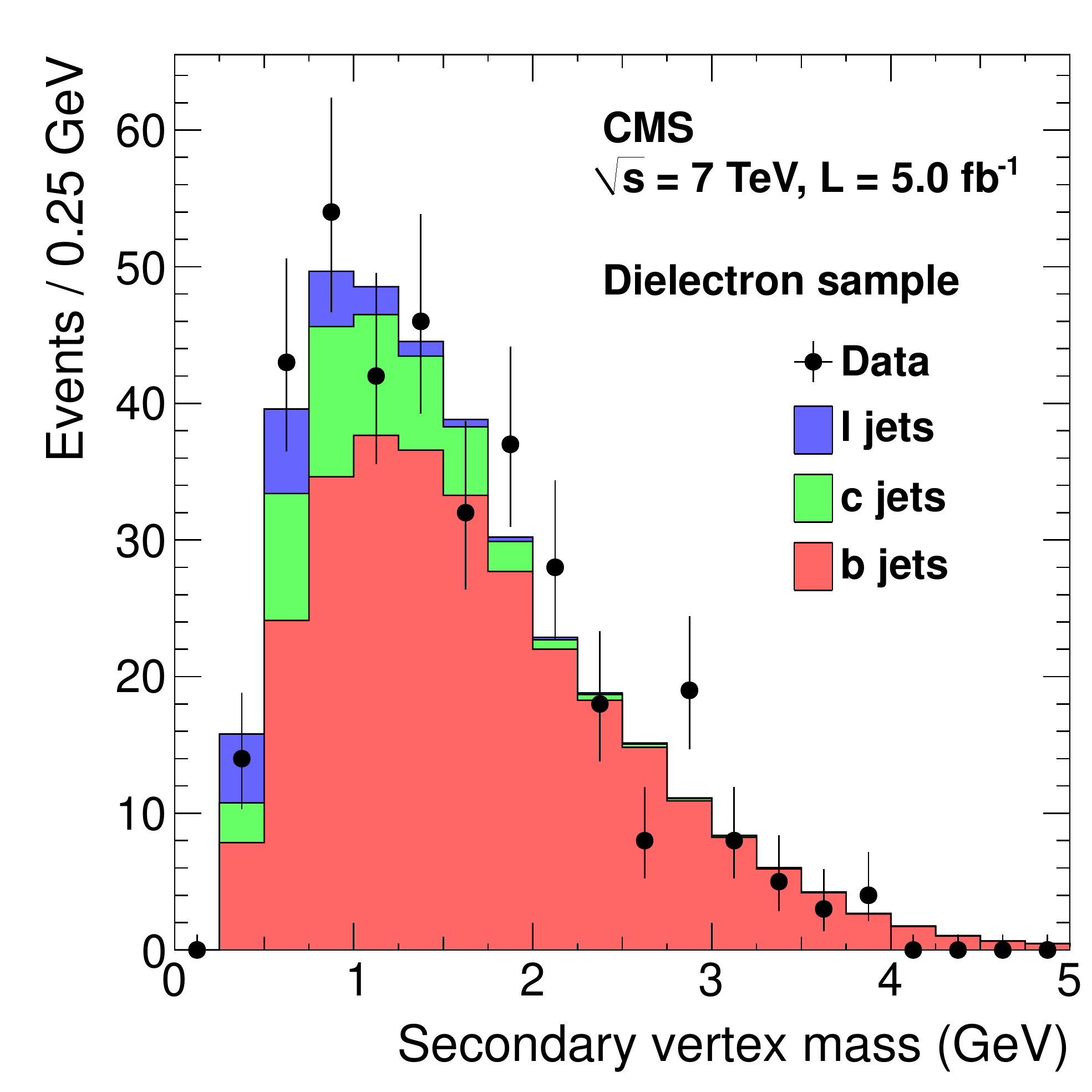}
    \caption{{
    Distributions of the secondary vertex mass of the leading (in \pt) b-tagged jet of the dimuon Z+2b-jets sample (left) and the subleading b-tagged jet of the dielectron Z+2b-jets sample (right).
    The overlaid distributions are the results of the fit described in the text.
      }
    }
    \label{fig:bpurFit_SimSum}
  \end{center}
\end{figure}

As described in detail in~\cite{Zbpaper},
templates are obtained from simulations to model the $M_\mathrm{SV}$ distributions for the various jet flavours;
separate templates are constructed for b jets, c jets, and light-parton jets.
These templates are used in maximum-likelihood fits to extract the fractions of b, c, and light-parton jets from the data for both the Z+1b-jet and the Z+2b-jets samples.
In the Z+2b-jets sample the distributions of the leading and subleading jets are fitted separately.
The results of fits to the one-dimensional
$M_\mathrm{SV}$ distributions after the Z+2b-jets selection are shown in Fig.~\ref{fig:bpurFit_SimSum}.
The fractions of correctly tagged b jets in the Z+1b-jet and Z+2b-jets samples are estimated to be ${\sim}55\%$ and 80--85\%, respectively.
The estimated fraction of correctly tagged b jets is checked by comparing the fit results to (i)
the results obtained with templates constructed from an independent MC sample,
and (ii) the direct expectations from simulation, and are found to be in agreement.
Effects due to gluon splitting in the modelling of the distributions have been studied and found to be negligible.

Subsequently, the fractions of correctly tagged b jets are transformed into the purities $P_\cPqb^\mathrm{Z+1b}$ and $P_\cPqb^\mathrm{Z+2b}$,
\ie the fractions of events in the two samples that contain correctly tagged b jets;
events in the Z+2b-jets sample with two correctly tagged b jets are considered as Z+2b-jets signal events,
whereas events with one mistagged jet in the Z+2b-jets sample are considered for the Z+1b-jet signal yield.
In order to estimate these ratios from the results of the one-dimensional fits,
the various combinations in which two jets are b-tagged in the Z+2b-jets sample are studied in simulations.
The systematic uncertainty related to the b purity is evaluated by varying the mistagging rates and production rates within their uncertainties.
As a cross-check, a fit is performed to
the two-dimensional distribution of the $M_\mathrm{SV}$ values
for the leading and subleading b-tagged jets,
and consistent results are obtained.

A small background from ZZ events is expected in the Z+2b-jets sample.
This contribution ($N_{\cPZ\cPZ}$) is estimated from MC simulations, using the cross section and uncertainty from the CMS measurement \cite{CMS_PAS_ZZ} for the normalization.
The yield from a SM Higgs boson with mass of 125\GeV~\cite{Higgs1,Higgs2,Higgs3} that decays into two b jets, and is produced in association with a Z boson,
is expected to be approximately $20\%$ of the ZZ contribution,
\ie 2.1 events in the Z($\mu\mu$)+2b-jets final state and 1.7 events in Z(ee)+2b-jets final state.
The resulting effect on the Z+2b-jets cross section is expected to be ${\sim}0.6\%$.

\begin{table}[t!h]
\centering
\topcaption{
The estimates of the purities, the \ttbar fractions, and the ZZ backgrounds for the various b-jet multiplicities and lepton flavours,
including statistical and systematic uncertainties.
}
\label{tab:backgrounds}
\begin{tabular}{ l c c c }
\hline
\rule[-7.0pt]{0pt}{18.5pt}%
Selection                  &  $P_\cPqb^\mathrm{Z+1b}$    & $f_{\ttbar}^\mathrm{Z+1b}$    & $N_{\cPZ\cPZ}^\mathrm{Z+1b}$  \\
\hline
Z($\mu\mu$)+1b-jet         &  $(53.7\pm1.1)\%$                &  $(5.2\pm0.7)\%$              & $73\pm24$        \\
Z(ee)+1b-jet               &  $(55.0\pm1.3)\%$                &  $(5.0\pm0.7)\%$              & $56\pm19$  \\
\hline
\rule[-7.0pt]{0pt}{18.5pt}%
                           &  $P_\cPqb^\mathrm{Z+2b}$    & $f_{\ttbar}^\mathrm{Z+2b}$    & $N_{\cPZ\cPZ}^\mathrm{Z+2b}$  \\
\hline
Z($\mu\mu$)+2b-jets        &  $(75.1\pm6.4)\%$                & $(13.0\pm1.9)\%$              & $12\pm4$  \\
Z(ee)+2b-jets              &  $(74.1\pm7.3)\%$                & $(14.0\pm2.3)\%$              &  $8\pm3$   \\
\hline
\end{tabular}
\end{table}

The background contributions are summarized in Table~\ref{tab:backgrounds}.
The backgrounds due to \ttbar and ZZ production increase when requiring two b-tagged jets,
because of the relatively harder spectra of these sources of background compared to the signal.
At the same time, the backgrounds due to light-parton jets decrease, since the probability of mistagging two jets is smaller.
The corrected signal yield ($N_\text{sig}$) is obtained by subtracting the backgrounds from the
number of selected events ($N_\text{rec}$),
and is estimated as
\begin{equation}\begin{split}\label{eq:bkgsubtract}
N_{\text{sig}}^\mathrm{Z+1b} &=
N^\mathrm{Z+1b}_{\text{rec}}\times(P_\cPqb^\mathrm{Z+1b}-f_{\ttbar}^\mathrm{Z+1b})
- N_{{\cPZ\cPZ}}^\mathrm{Z+1b} + f_{1\cPqb}^\mathrm{Z+2b}\times N^\mathrm{Z+2b}_{\text{rec}},\\
N_{\text{sig}}^\mathrm{Z+2b} &=
N^\mathrm{Z+2b}_{\text{rec}}\times(P_\cPqb^\mathrm{Z+2b}-f_{\ttbar}^\mathrm{Z+2b}) - N_{{\cPZ\cPZ}}^\mathrm{Z+2b}.
\end{split}\end{equation}
Here, $f_{1\cPqb}^\mathrm{Z+2b}$
is the fraction of events in the Z+2b-jets sample for which one jet is mistagged, which is $16\pm5\%$.
The resulting contribution to the Z+1b-jet cross section is ${\sim}1\%$.

\section{Efficiencies and migrations}
\label{sec:eff}

In order to extract a cross section at the particle level, the background-subtracted yields for the Z+1b-jet and the Z+2b-jets categories in Eq.~(\ref{eq:bkgsubtract})
are corrected for the
efficiencies in the selection of the dilepton pair and the b-tagged jets, as well as for the detector resolution effects.
Both the application of b tagging and jet reconstruction may induce migrations between the category of events containing one b jet
and that containing more than one, since the number of generated b jets and the number of correctly reconstructed b jets are, in general, not the same.
In order to estimate the cross sections for the different b-jet multiplicities,
the efficiency corrections (or `unfolding') are performed as a function of the number of b jets.

Particle-level b jets are defined by matching generated jets to a b hadron within $\DR<0.5$ of the jet axis.
No requirement is placed on the \pt of the hadron, and the generated jet is constructed from particle-level objects which include invisible particles.
The generated jets are clustered and selected with the same criteria used for the jets reconstructed in data.
Particle-level leptons are defined as `dressed' leptons, \ie adding to the lepton all generator-level photons within a cone of $\DR<0.1$.

The selection efficiency is factorised into two parts: the b-tagging efficiency ($\mathcal{E}_\cPqb$) and the
lepton selection efficiency ($\mathcal{E}_\ell$). The correction for the detector
resolution effects ($\mathcal{E}_\mathrm{r}$) is dominated by the jet energy resolution.
Finally, $\mathcal{E}_\mathrm{m}$ corrects for the efficiency loss associated with the selection criterion on the \MET significance in the Z+2b-jets event selection.

To account for migrations between different b-jet multiplicities, a $2\times2$ matrix equation is used.
Each efficiency factor is represented by a matrix (the matrices $\mathcal{E}_\ell$ and $\mathcal{E}_\mathrm{m}$ are diagonal).
The matrices are applied in an order reflecting the order of the selection requirements.
\begin{equation}
\renewcommand*{\arraystretch}{1.7}
\begin{pmatrix}
\sigma_{\mathrm{Z+1b}} \\
\sigma_{\mathrm{Z+2b}}
\end{pmatrix}
=
\frac{1}{\mathcal{L}}
\times
\mathcal{E}_\mathrm{r}^{-1}
\times
\mathcal{E}_{\ell}^{-1}
\times
\mathcal{E}_\cPqb^{-1}
\times
\mathcal{E}_\mathrm{m}^{-1}
\times
\begingroup
\renewcommand*{\arraystretch}{1.7}
\begin{pmatrix}
N_{\text{sig}}^{\mathrm{Z+1b}} \\
N_{\text{sig}}^{\mathrm{Z+2b}}
\end{pmatrix}
\endgroup\label{eq:master}.
\end{equation}
This equation is used to obtain the cross sections
for the production of a Z boson in association with exactly one b jet ($\sigma_{\mathrm{Z+1b}}$) or at least two b jets ($\sigma_{\mathrm{Z+2b}}$)
from the numbers of reconstructed signal events in the Z+1b-jet and Z+2b-jets categories.

\begin{figure}[ht!]
    \includegraphics[width=0.48\linewidth]{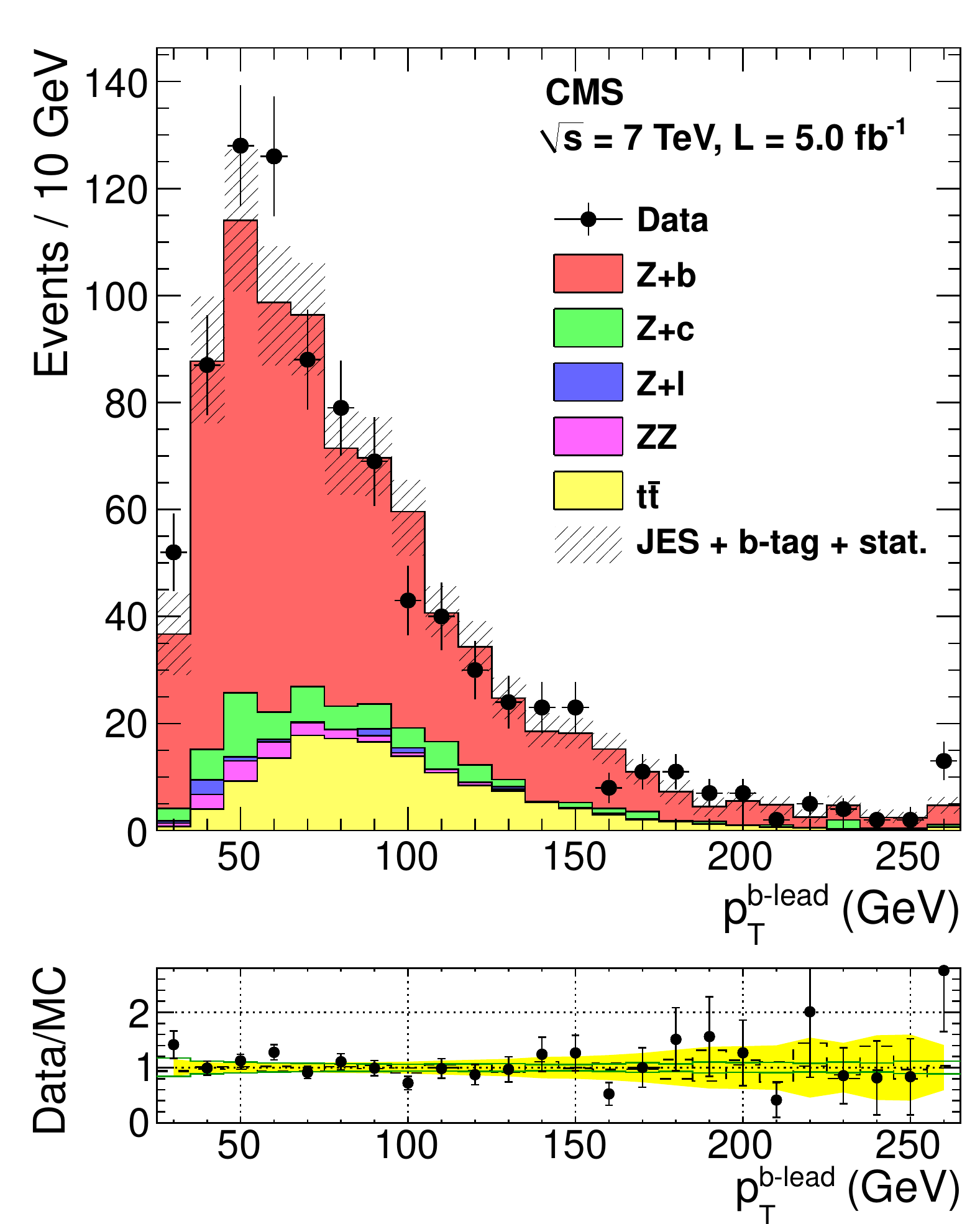}
    \includegraphics[width=0.48\linewidth]{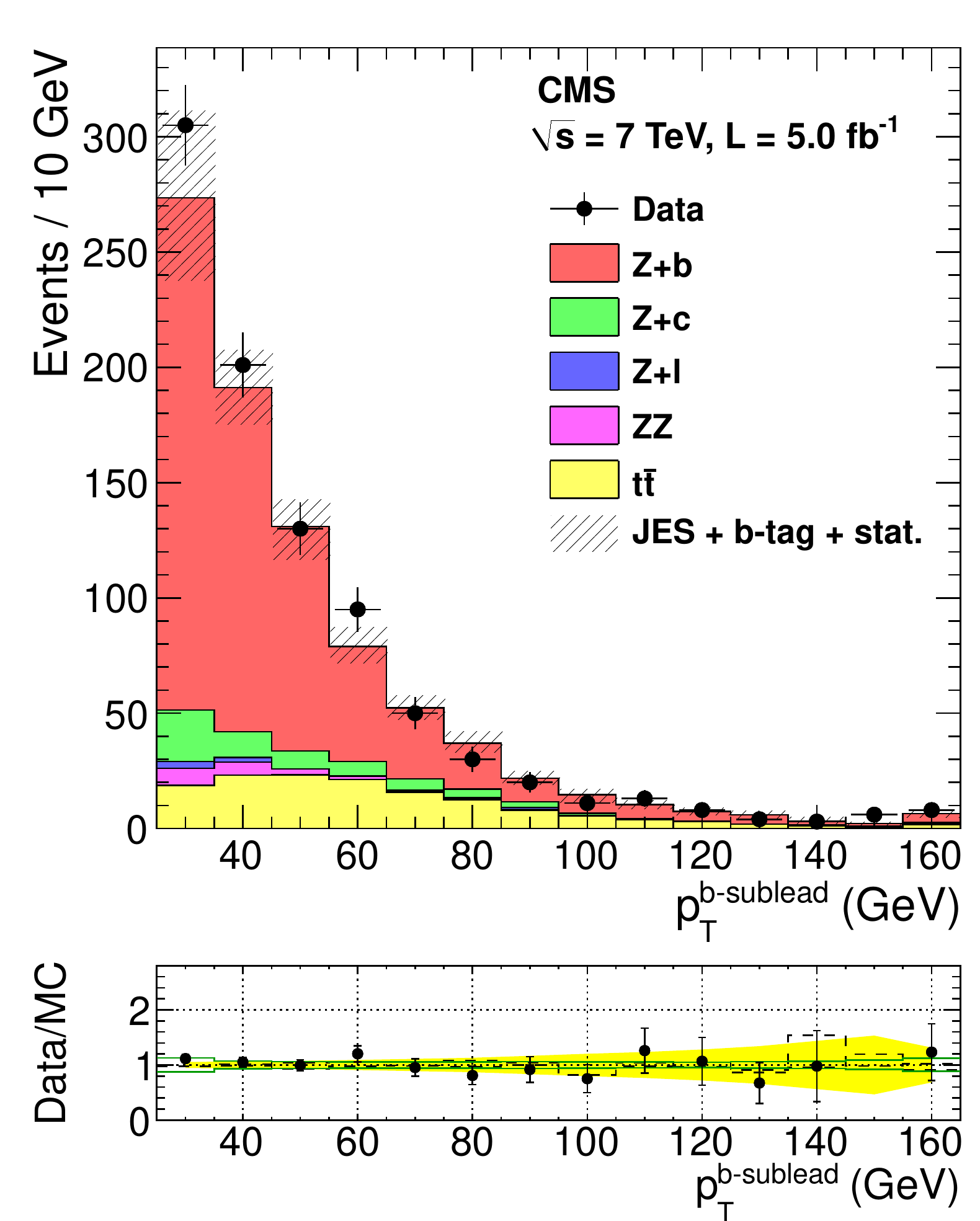}
   \caption{
     The combined muon+electron distributions of the \pt of the leading-\pt (left) and subleading-\pt (right) b-tagged jet
     for the Z+2b-jets sample.
     The simulated samples are normalized to the theoretical predictions.
     The last bin in both distributions contains the overflow, and the uncertainties in the simulations are shown as a hatched band.
     The data/simulation ratio shows the separate contributions to this uncertainty:
     the band represents the statistical uncertainty in the simulated yield,
     and the lines indicate the uncertainties related to the jet energy scale (dashed) and the b-tag scale factors (solid).
   }
    \label{fig:kinb_HEHESel}
\end{figure}

The MC signal sample is used to build the matrices,
with efficiencies from the simulation rescaled to match the
efficiencies observed in the data.
The \pt distributions for the leading (in \pt) and subleading b jets
after the Z+2b-jets selection are shown in Fig.~\ref{fig:kinb_HEHESel}.
The agreement between data and simulations in
Figs.~\ref{fig:met} and~\ref{fig:kinb_HEHESel} justifies the use of
this sample for the unfolding procedure.

The inclusive cross section for the production of a Z boson in association
with at least one b jet is the sum of the two cross sections in Eq.~(\ref{eq:master}), namely,
$\sigma_{\mathrm{Z+b}}\equiv\sigma_{\mathrm{Z+1b}}+\sigma_{\mathrm{Z+2b}}$.
The ratio of this cross section to the cross section for the production of a Z boson with any kind of jet
is denoted $\sigma_\mathrm{Z+b/Z+j}$.
The cross sections are defined using the same acceptance for the different lepton flavours: events have
leptons with $\pt^{\ell}>20$\GeV and $|\eta^\ell|<2.4$, a dilepton invariant mass $76<M_{\ell\ell}<106$\GeV,
and jets with $\pt^{j}>25$\GeV and $\abs{\eta^j}<2.1$, and a separation between the leptons and the jets of $\DR(\ell,j)>0.5$.

The terms in Eq.~(\ref{eq:master})
related to the b-tagging and \MET efficiencies are found to be very similar for the
muon and the electron channels, as expected. For the lepton selection
efficiencies, results are found to be almost identical between the two
b-jet multiplicity bins, which is expected since the requirement of
$\DR(\ell,j)>0.5$ effectively renders the lepton selection
insensitive to the jet multiplicity.

\section{Systematic uncertainties}
\label{sec:syst}

The following sources of systematic uncertainties are considered:
\begin{itemize}
\item Background from light-parton jets: For the estimate of the background due to mistagged b jets,
the main source of uncertainty arises from the fit uncertainty in the fraction of b jets in the Z+1b-jet and Z+2b-jets samples.
Another source of uncertainty originates from the ambiguity when estimating the number of events containing zero, one, or two b jets in the Z+2b-jets sample.
The corresponding systematic uncertainty is estimated by varying the (mis)tagging efficiencies according to their uncertainties.
Studies show that no significant differences are observed when comparing the $M_\mathrm{SV}$ templates obtained with different MC generators,
and that the template acquired from simulation correctly describes the distribution observed in data~\cite{CMS_PAS_BTV-11-002,CMS_PAS_BTV-13-001}.
\item Background from \ttbar: The main source of uncertainty in the estimate of the \ttbar background is the statistical uncertainty from the fit.
An additional uncertainty originates from the modelling of the signal and background shapes.
The probability density functions used in the estimate of the \ttbar background are obtained in three distinct ways:
with templates based on simulation and on data,
and by modelling the contributions with an empirical parameterization.
The systematic uncertainty is estimated from the differences between the three methods.
\item The ZZ background: The uncertainty in the overall normalization
is taken from the CMS measurement~\cite{CMS_PAS_ZZ}.
Correlated sources of uncertainties (such as the luminosity) are ignored to avoid double counting.
\end{itemize}
All background-related systematic uncertainties are listed in Table~\ref{tab:backgrounds}, and are
  propagated to the cross section estimate following Eq.~(\ref{eq:bkgsubtract}).
Other systematic uncertainties, estimated via Eq.~(\ref{eq:master}), are:
\begin{itemize}
\item The b-tagging efficiency and the mistagging fraction:
The uncertainties of the b-tagging efficiencies and mistagging fractions are estimated in the
data as functions of the \pt and $\eta$ of the jet, combining the
various methods discussed in Ref.~\cite{CMS_PAS_BTV-11-002}.
  These uncertainties affect the b-tagging efficiencies as described in Section~\ref{sec:eff}.
  The $\pt$-dependent uncertainties in the jet tagging efficiency, 3--8\% for $\pt>30\GeV$ and 12\% for $\pt<30\GeV$,
  are propagated to the b-tagging data/simulation scale factors, by varying these according to the corresponding uncertainties with the flavor of each jet.
  The uncertainty in the mistagging fraction, which enters the calculation of the
  event weight at second order, is found to have a negligible
  impact.
\item Jet energy scale (JES) and resolution (JER):
  The jet energy calibration is based on MC simulations, while residual corrections are used to account for the small differences between data and simulation.
  The JES uncertainty is taken from Ref.~\cite{CMS_PAS_JME-10-001} and amounts to
  3--5\% depending on the \pt and $\eta$ of the jets.
  The JER uncertainty is taken to be 10\%, after
  degrading the simulated resolution by 10\% to match that measured in the
  data. Both affect $\mathcal{E}_\mathrm{r}$.
  Studies of simulated samples show that these JES corrections are good for jets from bottom quarks.
\item Effect from pileup: The total inelastic cross section used to infer
  the pileup in data from the instantaneous luminosity is varied by $\pm$5\%,
  thereby affecting the pileup distribution in the simulated samples and covering the uncertainties due to pileup modelling.
  It is then propagated to the estimation of the unfolding matrices where
  it affects mainly the lepton efficiency factors through the lepton isolation
  requirements.
\item Requirement on \MET: The requirement on the \MET significance removes ${\sim}$2\% of the
  Z+2b-jets signal contribution, which is evaluated from simulation.
  The systematic uncertainty is estimated by varying each component entering the \MET calculation
  within its uncertainty. This includes contributions from JES and JER as discussed above,
  unclustered energy (10\%), $\tau$ leptons (3\%), electrons and photons (0.6--1.5\%), and muons (0.2\%)~\cite{CMS_PAS_JME-10-009}.
\item MC statistics: While the MC statistics suffice for the Z+1b-jet sample,
  they lead to uncertainties of several percent in correction
  factors involving the Z+2b-jets sample.
\item Luminosity: The uncertainty of the integrated luminosity recorded by CMS is 2.2\% in the 2011 data set~\cite{CMSLumi}.
\item Dilepton selection efficiencies:
The systematic uncertainty of the scale factor per lepton, which is applied to simulated events to compensate for data/simulation differences,
is obtained with the tag-and-probe method, and is less than 0.4\% for muons and 1.0\% for electrons.
\item Theory: The effect of uncertainties in the renormalization and factorization scales is estimated using \MCFM~\cite{Maltoni:2005wd}.
The impact of scale variations on the \pt of the b jets is used in the unfolding procedure to estimate the effect on the cross section.
Similarly, the \pt of the dilepton pair is varied according to the difference observed between data and simulation to estimate the impact on the unfolding.

Furthermore, the effect due to MPIs on the acceptance of Z+b-jets events is studied by artificially reducing their contribution by a factor two. 
This is done by applying a veto on the azimuthal angle $\Delta\phi_{\cPZ,\cPqb\cPqb}$, which has been shown to be a discriminant observable for MPIs~\cite{MPIFSQ}.
The effect of this requirement on the cross sections has been found to be less than 0.5\%.

Together, this leads to an uncertainty of at most 3\% in the cross sections.
\item Vertex association: For the estimate of the cross section ratio $\sigma_\mathrm{Z+b/Z+j}$,
an additional uncertainty arises from the contribution of jets not associated with the primary vertex.
After the requirement on the momentum fraction of tracks originating from the primary vertex,
the background due to pileup is estimated from simulation to be 2.2\%.
The efficiency of the requirement is estimated from the distribution of this observable in data before applying the requirement.
The corresponding systematic
uncertainty is evaluated by comparing the distributions of this observable in data and simulation; it is assumed that the difference observed for
the variable used for the vertex association is entirely due to events originating from pileup.
This assumption results in a systematic uncertainty of 18\%
in the pileup contamination, fully correlated between the electron and muon channels.
\end{itemize}

\begin{table}[t!h]
\centering
  \caption{ \label{tab:uncert} {Fractional uncertainties in the measured cross sections, grouped according to the correlation between the channels. }}
\label{tab:syst}
\begin{tabular}{ l c c c c c }\hline
                                 & \multicolumn{2}{c }{$\mu\mu$ $(\%)$} & \multicolumn{2}{c }{ee (\%)} \\
\hline
                                 &  Z+1b   &  Z+2b                      &  Z+1b   &  Z+2b                  \\
\hline
 Uncorrelated                    &         &                            &         &                        \\
\hline
 b purity                        &   3.0   &  12.7                      &   3.3   &  15.1                  \\
 \ttbar                          &   1.7   &   3.8                      &   1.7   &   4.8                  \\
 Dilepton selection              &   1.0   &   1.0                      &   2.0   &   2.0                  \\
 MC statistics                   &   0.9   &   4.2                      &   1.2   &   5.1                  \\
\hline
 Correlated                      &         &                            &         &                        \\
\hline
 b-tagging efficiency            &   3.6   &   9.0                      &   3.6   &   9.0                  \\
 Jet energy scale                &   2.0   &   3.6                      &   2.0   &   3.6                  \\
 Theory                          &   1.8   &   3.0                      &   1.8   &   3.0                  \\
 Luminosity                      &   2.2   &   2.2                      &   2.2   &   2.2                  \\
 ZZ                              &   0.4   &   1.2                      &   0.5   &   1.4                  \\
 Jet energy resolution           &   0.6   &   0.7                      &   0.6   &   0.7                  \\
 Pileup                          &   0.3   &   0.3                      &   0.3   &   0.3                  \\
 Mistag                          &   0.0   &   0.1                      &   0.0   &   0.1                  \\
\hline
 Total stat. uncertainty         &   0.9   &   4.5                      &   1.0   &   5.4                  \\
 Total syst. uncertainty         &   6.3   &  17.4                      &   6.7   &  19.8                  \\
\hline
\end{tabular}
\end{table}

The systematic uncertainties are summarized in Table~\ref{tab:uncert}.
The uncertainties are presented separately for the muon and electron channels,
and for the Z+1b-jet and Z+2b-jets measurements.

\section{Kinematic observables}

One of the observables of interest for searches in the Z+2b-jets final state is the invariant mass of the b-jet pair ($M_{\cPqb\cPqb}$).
This observable is, for example, used in the study of the Higgs boson produced in association with a Z boson and decaying into two b jets,
in the Z($\ell\ell$)H(bb) final state~\cite{Hbb,Hbb2}.
Other kinematic observables in the Z+2b-jets final state relevant to searches for undiscovered processes
are the transverse momentum of the dilepton ($\pt^\cPZ$) and the dijet ($\pt^{\cPqb\cPqb}$) pair,
and the angle between the dilepton pair and the dijet pair ($\Delta\phi_{\cPZ,\cPqb\cPqb}$).
The distributions of these observables are compared with the predictions from \MADGRAPH,
including uncertainties due to the jet energy scale and the b-tagging efficiencies, as well as the uncertainties due to limited MC statistics.
More than two jets are b-tagged in less than 2\% of the Z+2b-jets events, and in this case the two highest-$\pt$ jets are considered.

\begin{figure}[ht!p]
\centering
    \includegraphics[width=0.48\linewidth]{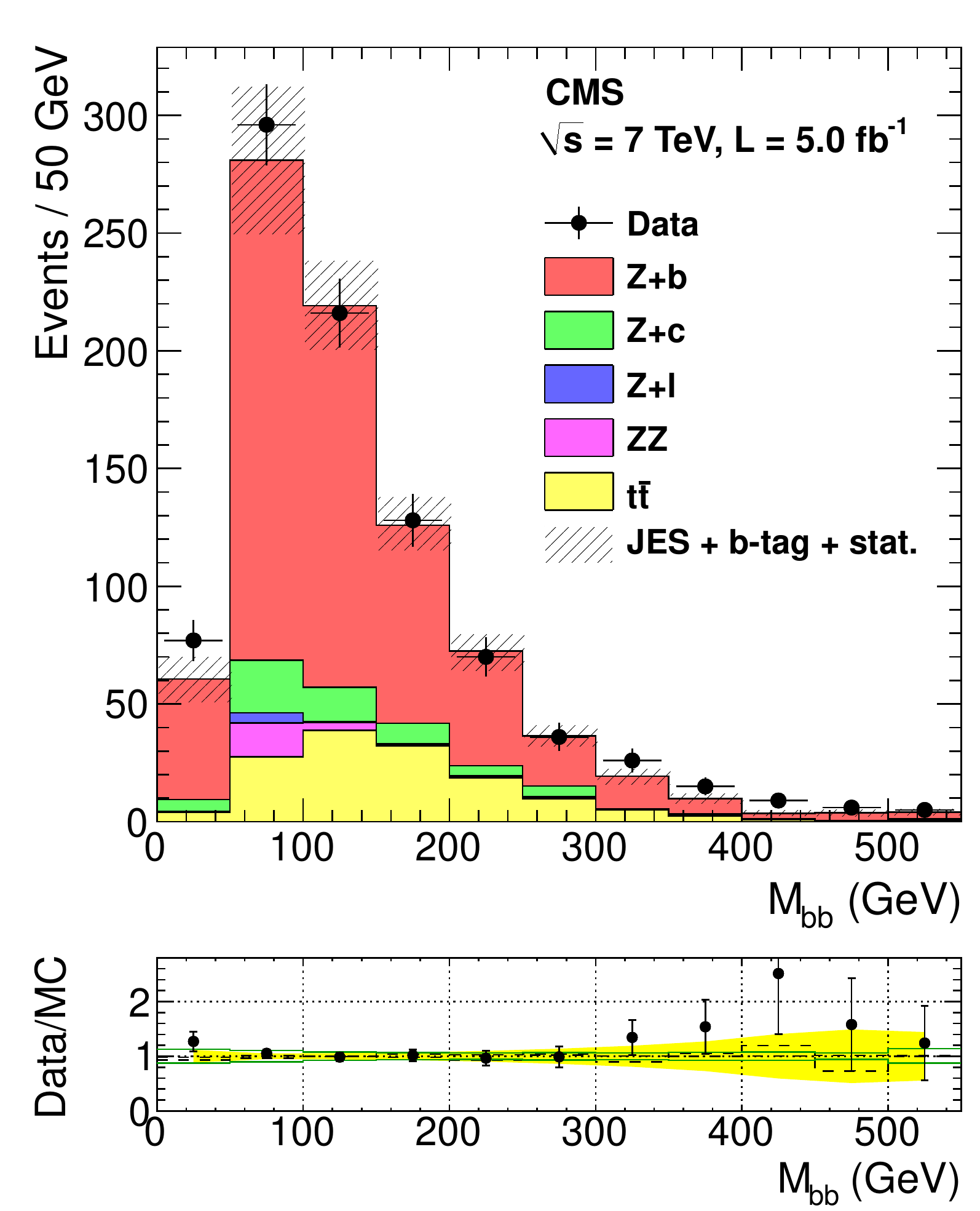}
    \includegraphics[width=0.48\linewidth]{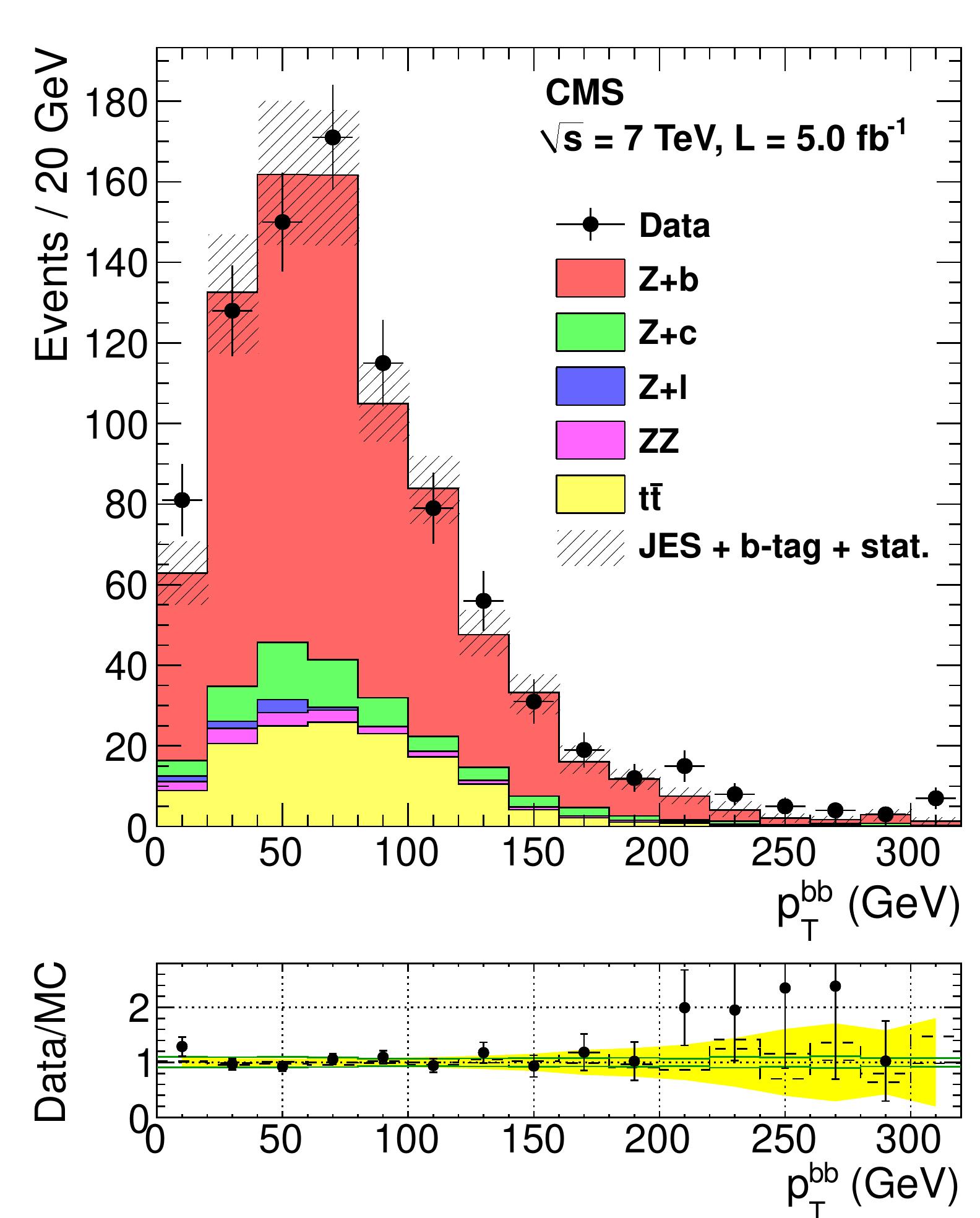}
    \includegraphics[width=0.48\linewidth]{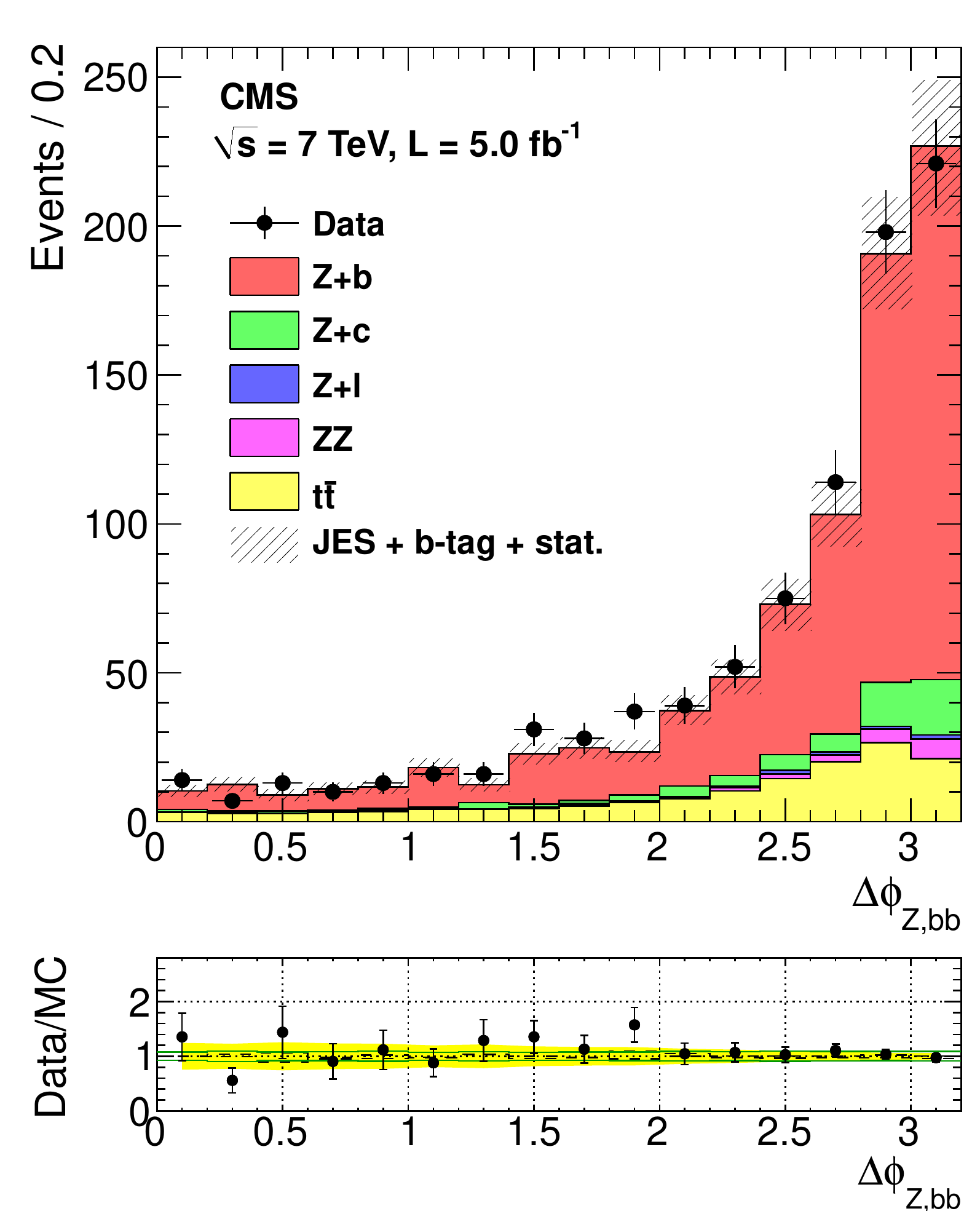}
    \includegraphics[width=0.48\linewidth]{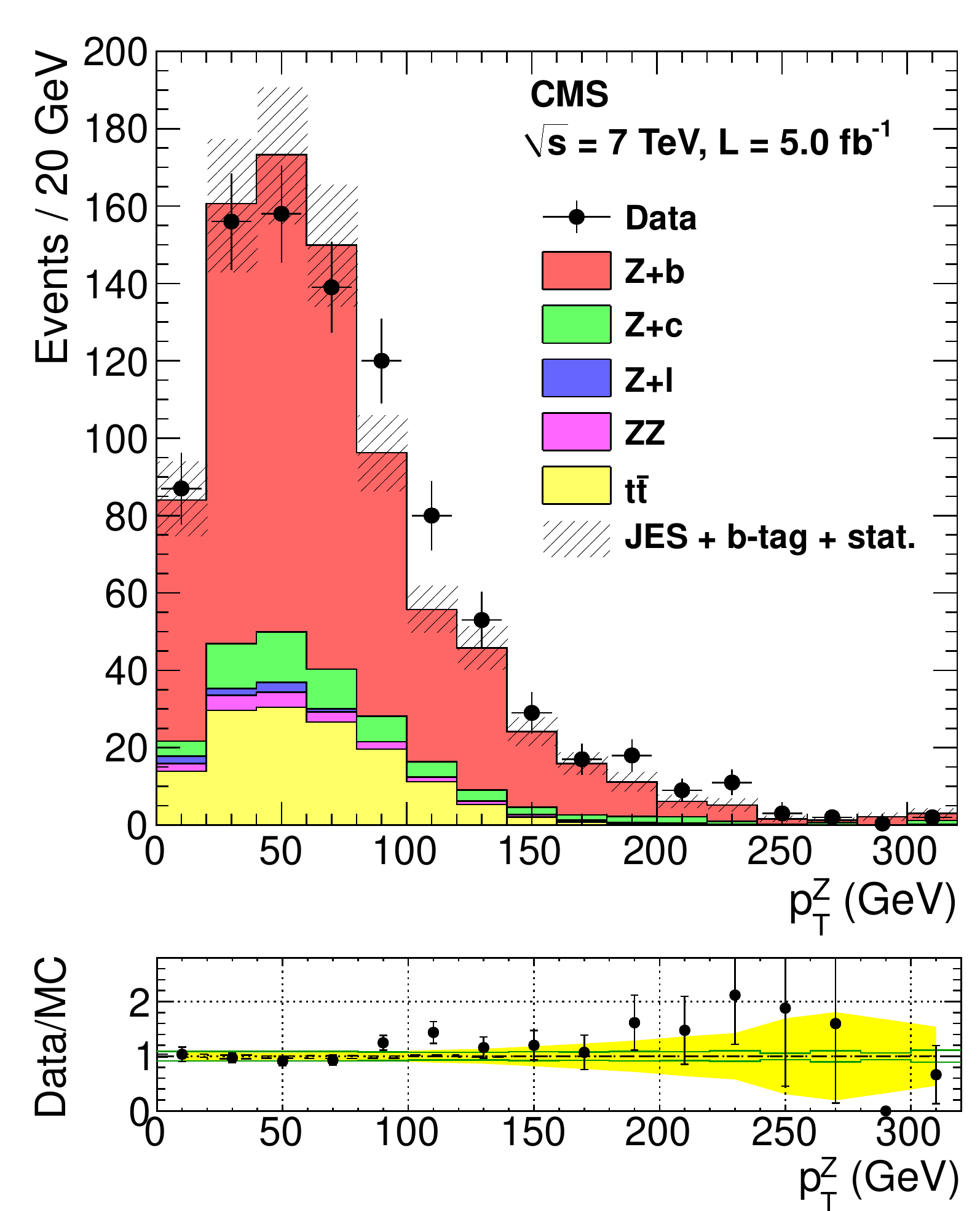}
    \caption{
     Distributions of kinematic observables for the Z+2b-jets selection of the combined electron and muon samples,
     and a comparison with the simulated samples that are normalized to the theoretical predictions.
     Top left: the dijet mass of the two b-tagged jets.
     Top right: the $\pt$ distribution of the dijet pair.
     Left bottom: the azimuthal angle $\phi$ between the Z boson and the dijet system.
     Right bottom: the $\pt$ distribution of the dilepton pair.
     The right-most bin in the last three plots contains the overflow. Uncertainties in the predictions are shown as a hatched band.
     The data/simulation ratio shows the separate contributions to this uncertainty:
     the band represents the statistical uncertainty on the simulated yield,
     and the lines indicate the uncertainties related to the jet energy scale (dashed) and the b-tagging scale factors (solid).
   }
    \label{fig:kinZ2_HEHEMETSel}
\end{figure}

The distributions of $M_{\cPqb\cPqb}$ and $\pt^{\cPqb\cPqb}$,
presented in Fig.~\ref{fig:kinZ2_HEHEMETSel} (top left and top right, respectively),
show agreement with the predictions.
The excess of data in the overflow bin at high values of $\pt^{\cPqb\cPqb}$ is not concentrated in any particular region.
The distribution of $\Delta\phi_{\cPZ,\cPqb\cPqb}$, shown in Fig.~\ref{fig:kinZ2_HEHEMETSel} (bottom left),
shows agreement with the predictions as well, both in the collinear and back-to-back regions.
This is especially relevant with respect to contributions from MPIs,
which are expected to have less correlated kinematics than those from the Z+2b-jets process,
and will therefore give a uniform distribution in $\Delta\phi_{\cPZ,\cPqb\cPqb}$.

On the other hand, the $\pt^\cPZ$ distribution shows a harder spectrum in data than predicted,
as shown in Fig.~\ref{fig:kinZ2_HEHEMETSel} (right bottom).
An overall excess of events is observed for $\pt^\cPZ>80\GeV$,
in particular in the region around 100\GeV.
This trend is consistent with the earlier CMS publication~\cite{Zbpaper},
where a similar discrepancy is observed for the $\pt^\cPZ$ observable in the Z+b-jets final state.
A harder spectrum for the $\pt^\cPZ$ observable is predicted in four-flavour calculations with massive b quarks at NLO~\cite{Frederix:2011qg},
which might explain the observed disagreement.

The effect of the disagreement on the estimate of the cross sections has been studied and is included in the systematic uncertainties,
as described in Section~\ref{sec:syst}.
Furthermore, a bin-by-bin reweighting of the predictions according to the observed discrepancy in the $\pt^\cPZ$ observable has been performed,
and this improves the agreement in other observables where differences are observed.

\section{Cross sections}

The cross sections are estimated per b-jet multiplicity bin and for each lepton flavour separately.
The results are summarized in Table~\ref{tab:xsec_elmu}.

\begin{table}[t!h]
\centering
  \topcaption{
    Cross sections at the particle level for the production of a Z boson with exactly one b~jet, with at least two b~jets, and with
    at least one b~jet, and the ratio with respect to the production of a Z boson in association with at least one jet of any flavour.
    The first uncertainty is statistical, and the second systematic.
  }
\label{tab:xsec_elmu}
\begin{tabular}{ l c c }
\hline
Cross section                        & $\mu\mu$                   & ee                        \\
\hline
    $\sigma_\mathrm{Z+1b}$ (pb)      & $ 3.52 \pm 0.03 \pm 0.22$  & $ 3.51 \pm 0.04 \pm 0.23$ \\
    $\sigma_\mathrm{Z+2b}$ (pb)      & $ 0.38 \pm 0.02 \pm 0.07$  & $ 0.32 \pm 0.02 \pm 0.06$ \\
    $\sigma_\mathrm{Z+b}$ (pb)       & $ 3.91 \pm 0.04 \pm 0.23$  & $ 3.84 \pm 0.04 \pm 0.24$ \\
    \hline
    $\sigma_\mathrm{Z+b/Z+j}$ ($\%$) & $ 5.23 \pm 0.04 \pm 0.24 $ & $ 5.08 \pm 0.05 \pm 0.24$ \\
\hline
\end{tabular}
\end{table}

Using the best linear unbiased estimator~\cite{BLUE}, results for the $\mu\mu$ and ee channels are
found to be consistent with a $\chi^2$ probability of 42\% for the Z+1b and 78\% for the Z+2b cases.
They are therefore combined into a single measurement using the optimal set of coefficients that minimise
the total uncertainty in the combined result, taking into account statistical and systematic uncertainties
and their correlations. The results are summarized in Table~\ref{tab:xsec_lep} and are then compared with various predictions.

The expectations from {\MADGRAPH}, in both the 5F and the 4F schemes,
are estimated using a global $K$ factor to correct the inclusive Drell--Yan cross section for next-to-NLO effects~\cite{DY_NNLO}.
The expectations from a\MCATNLO, at NLO, are also estimated using both 5F calculations and 4F calculations with massive b quarks~\cite{Frederix:2011qg}.
The events simulated with \MADGRAPH and a\MCATNLO are interfaced with the \PYTHIA parton shower simulation.
The settings used for the predictions from \MADGRAPH and a{\MCATNLO} are described in detail in~\cite{ewk11015}.

The NLO prediction from \MCFM is at the parton level.
The \MCFM calculations are estimated with the CTEQ6mE PDF, and the renormalization and factorization scales are set to the invariant mass of the dilepton pair.

\begin{table}[th!]
\centering
  \topcaption{
    Cross sections for the production of a Z boson
    with exactly one b~jet, with at least two b~jets, with
    at least one b~jet, and the ratio with respect to at least one jet of any flavour, showing the statistical and systematic uncertainties.
    The expectations from \MADGRAPH, \MCFM and a\MCATNLO include uncertainties due to scale variations.
  }
\label{tab:xsec_lep}
\resizebox{\textwidth}{!}{
\begin{tabular}{ l | c | c c c c c }
\hline
Cross section                        & Measured             & \MADGRAPH     & a\MCATNLO          & \MCFM             & \MADGRAPH              & a\MCATNLO          \\
                                     &                      & (5F)          & (5F)                   & (parton level)         & (4F)                   & (4F)                   \\
\hline
\rule[-7.0pt]{0pt}{18.5pt}%
    $\sigma_\mathrm{Z+1b}$ (pb)      & $3.52\pm0.02\pm0.20$ & $3.66\pm0.22$ & $3.70^{+0.23}_{-0.26}$ & $3.03^{+0.30}_{-0.36}$ & $3.11^{+0.47}_{-0.81}$ & $2.36^{+0.47}_{-0.37}$ \\
    $\sigma_\mathrm{Z+2b}$ (pb)      & $0.36\pm0.01\pm0.07$ & $0.37\pm0.07$ & $0.29^{+0.04}_{-0.04}$ & $0.29^{+0.04}_{-0.04}$ & $0.38^{+0.06}_{-0.10}$ & $0.35^{+0.08}_{-0.06}$ \\
\rule[-7.0pt]{0pt}{18.5pt}%
    $\sigma_\mathrm{Z+b}$  (pb)      & $3.88\pm0.02\pm0.22$ & $4.03\pm0.24$ & $3.99^{+0.25}_{-0.29}$ & $3.23^{+0.34}_{-0.40}$ & $3.49^{+0.52}_{-0.91}$ & $2.71^{+0.52}_{-0.41}$ \\
    \hline
\rule[-7.0pt]{0pt}{18.5pt}%
    $\sigma_\mathrm{Z+b/Z+j}$ ($\%$) & $5.15\pm0.03\pm0.25$ & $5.35\pm0.11$ & $5.38^{+0.34}_{-0.39}$ & $4.75^{+0.24}_{-0.27}$ & $4.63^{+0.69}_{-1.21}$ & $3.65^{+0.70}_{-0.55}$ \\
\hline
\end{tabular}
}
\end{table}

Uncertainties in the theoretical predictions are estimated by varying
the renormalization and factorization scales by a factor two up and down.
For the \MADGRAPH 5F prediction, the scales are varied in a correlated manner,
whereas the scales are varied in an uncorrelated way for the other predictions, which leads to a
larger estimate for the uncertainty.
The uncertainties in the 4F predictions amount to 15--20\%, as expected~\cite{ZbbNLOunc}.
Variations of the PDFs (using MSTW2008~\cite{MSTW}, CTEQ6, and CT10~\cite{CT10} PDF sets), jet matching scale (up to a factor of two),
and mass of the b quark (between 4.4 and 5.0\GeV) all result in smaller uncertainties.
A more detailed description of the methods to estimate these uncertainties is given in~\cite{ewk11015}.

The measured cross sections are consistent, within uncertainties,
with the expectations in the 5F scheme from both \MADGRAPH and a\MCATNLO.
Compared to the predictions from \MADGRAPH and a\MCATNLO in the 5F scheme, the predictions from \MCFM are approximately $20\%$ lower.
The predictions by \MADGRAPH and a\MCATNLO from calculations in the 4F scheme, compared to the predictions in the 5F scheme,
show a reduction of the Z+1b-jet production rate, when the other b jet in the final state is produced outside of the acceptance.

A difference of approximately two standard deviations
is observed when comparing to the parton-level prediction from \MCFM for the Z+b-jets cross section.
Since the correction factor from parton level to hadron level is smaller than one~\cite{Zbpaper},
this difference is not explained by hadronization effects.
The difference remains when measuring the cross section ratio,
which excludes an explanation based on experimental systematic effects that are shared between the Z+jets and the Z+b-jets final states,
such as luminosity, and the reconstruction of jets and leptons.
These results indicate that the difference observed with \MCFM is specific to the modelling of the Z+b-jets final state.

The largest discrepancy is observed when comparing the measured Z+1b-jet cross section with the predictions in the 4F scheme.
In particular, the prediction from a\MCATNLO in the 4F scheme shows a discrepancy of more than two standard deviations compared to the measurement.

\section{Conclusions}

The production of Z($\ell\ell$)+b-jets, with $\ell\ell$ = $\mu\mu$ or ee,
has been studied for events containing leptons with $\pt^{\ell} > 20\GeV$, $\abs{\eta^\ell}<2.4$,
a dilepton invariant mass $76< M_{\ell\ell} < 106\GeV$, jets with $\pt^j >25\GeV$ and $\abs{\eta^j} < 2.1$, and a separation between the leptons and
the jets of $\Delta{R}(\ell,j) > 0.5$.
The Z+b-jets cross sections have been measured,
at the level of stable final-state particles, for a Z boson produced with exactly one or at least two b jets.
In addition, a cross section ratio has been extracted for a \cPZ\ boson produced with at least one \cPqb\ jet relative to
a \cPZ\ boson produced with at least one jet.

The cross section measurements are in agreement with the expectations from \MADGRAPH and a\MCATNLO in the five-flavour scheme.
A difference of approximately two standard deviations is observed when comparing the cross sections with the predictions from \MCFM at the parton level,
and the comparison with the cross section ratio indicates that the difference is specific to the modelling of the Z+b-jets final state.
Comparisons with the predictions in the four-flavour scheme, in particular from a\MCATNLO, show a disagreement of more than two standard deviations
in the Z+1b-jet final state.

Comparisons of the kinematic properties of Z+2b-jets production with the predictions from \MADGRAPH
in the five-flavour scheme show potential limitations of the existing MC event generators
that employ the matrix element plus parton shower approach at leading order with massless b quarks.
While these observations should be confirmed with more data, next-to-leading-order simulations and/or simulations with massive quarks
could possibly provide a better description of the data in certain regions of phase space.

\section*{Acknowledgements}

We thank Marco Zaro for his help in simulating a{\sc MC@NLO} samples for this analysis.

\hyphenation{Bundes-ministerium Forschungs-gemeinschaft Forschungs-zentren} We congratulate our colleagues in the CERN accelerator departments for the excellent performance of the LHC and thank the technical and administrative staffs at CERN and at other CMS institutes for their contributions to the success of the CMS effort. In addition, we gratefully acknowledge the computing centres and personnel of the Worldwide LHC Computing Grid for delivering so effectively the computing infrastructure essential to our analyses. Finally, we acknowledge the enduring support for the construction and operation of the LHC and the CMS detector provided by the following funding agencies: the Austrian Federal Ministry of Science and Research and the Austrian Science Fund; the Belgian Fonds de la Recherche Scientifique, and Fonds voor Wetenschappelijk Onderzoek; the Brazilian Funding Agencies (CNPq, CAPES, FAPERJ, and FAPESP); the Bulgarian Ministry of Education and Science; CERN; the Chinese Academy of Sciences, Ministry of Science and Technology, and National Natural Science Foundation of China; the Colombian Funding Agency (COLCIENCIAS); the Croatian Ministry of Science, Education and Sport, and the Croatian Science Foundation; the Research Promotion Foundation, Cyprus; the Ministry of Education and Research, Recurrent financing contract SF0690030s09 and European Regional Development Fund, Estonia; the Academy of Finland, Finnish Ministry of Education and Culture, and Helsinki Institute of Physics; the Institut National de Physique Nucl\'eaire et de Physique des Particules~/~CNRS, and Commissariat \`a l'\'Energie Atomique et aux \'Energies Alternatives~/~CEA, France; the Bundesministerium f\"ur Bildung und Forschung, Deutsche Forschungsgemeinschaft, and Helmholtz-Gemeinschaft Deutscher Forschungszentren, Germany; the General Secretariat for Research and Technology, Greece; the National Scientific Research Foundation, and National Innovation Office, Hungary; the Department of Atomic Energy and the Department of Science and Technology, India; the Institute for Studies in Theoretical Physics and Mathematics, Iran; the Science Foundation, Ireland; the Istituto Nazionale di Fisica Nucleare, Italy; the Korean Ministry of Education, Science and Technology and the World Class University program of NRF, Republic of Korea; the Lithuanian Academy of Sciences; the Ministry of Education, and University of Malaya (Malaysia); the Mexican Funding Agencies (CINVESTAV, CONACYT, SEP, and UASLP-FAI); the Ministry of Business, Innovation and Employment, New Zealand; the Pakistan Atomic Energy Commission; the Ministry of Science and Higher Education and the National Science Centre, Poland; the Funda\c{c}\~ao para a Ci\^encia e a Tecnologia, Portugal; JINR, Dubna; the Ministry of Education and Science of the Russian Federation, the Federal Agency of Atomic Energy of the Russian Federation, Russian Academy of Sciences, and the Russian Foundation for Basic Research; the Ministry of Education, Science and Technological Development of Serbia; the Secretar\'{\i}a de Estado de Investigaci\'on, Desarrollo e Innovaci\'on and Programa Consolider-Ingenio 2010, Spain; the Swiss Funding Agencies (ETH Board, ETH Zurich, PSI, SNF, UniZH, Canton Zurich, and SER); the National Science Council, Taipei; the Thailand Center of Excellence in Physics, the Institute for the Promotion of Teaching Science and Technology of Thailand, Special Task Force for Activating Research and the National Science and Technology Development Agency of Thailand; the Scientific and Technical Research Council of Turkey, and Turkish Atomic Energy Authority; the Science and Technology Facilities Council, UK; the US Department of Energy, and the US National Science Foundation.

Individuals have received support from the Marie-Curie programme and the European Research Council and EPLANET (European Union); the Leventis Foundation; the A. P. Sloan Foundation; the Alexander von Humboldt Foundation; the Belgian Federal Science Policy Office; the Fonds pour la Formation \`a la Recherche dans l'Industrie et dans l'Agriculture (FRIA-Belgium); the Agentschap voor Innovatie door Wetenschap en Technologie (IWT-Belgium); the Ministry of Education, Youth and Sports (MEYS) of Czech Republic; the Council of Science and Industrial Research, India; the Compagnia di San Paolo (Torino); the HOMING PLUS programme of Foundation for Polish Science, cofinanced by EU, Regional Development Fund; and the Thalis and Aristeia programmes cofinanced by EU-ESF and the Greek NSRF.
\bibliography{auto_generated}   

\cleardoublepage \appendix\section{The CMS Collaboration \label{app:collab}}\begin{sloppypar}\hyphenpenalty=5000\widowpenalty=500\clubpenalty=5000\textbf{Yerevan Physics Institute,  Yerevan,  Armenia}\\*[0pt]
S.~Chatrchyan, V.~Khachatryan, A.M.~Sirunyan, A.~Tumasyan
\vskip\cmsinstskip
\textbf{Institut f\"{u}r Hochenergiephysik der OeAW,  Wien,  Austria}\\*[0pt]
W.~Adam, T.~Bergauer, M.~Dragicevic, J.~Er\"{o}, C.~Fabjan\cmsAuthorMark{1}, M.~Friedl, R.~Fr\"{u}hwirth\cmsAuthorMark{1}, V.M.~Ghete, N.~H\"{o}rmann, J.~Hrubec, M.~Jeitler\cmsAuthorMark{1}, W.~Kiesenhofer, V.~Kn\"{u}nz, M.~Krammer\cmsAuthorMark{1}, I.~Kr\"{a}tschmer, D.~Liko, I.~Mikulec, D.~Rabady\cmsAuthorMark{2}, B.~Rahbaran, C.~Rohringer, H.~Rohringer, R.~Sch\"{o}fbeck, J.~Strauss, A.~Taurok, W.~Treberer-Treberspurg, W.~Waltenberger, C.-E.~Wulz\cmsAuthorMark{1}
% ...
\vskip\cmsinstskip
% ...
1:~~Also at Vienna University of Technology, Vienna, Austria\\
2:~~Also at CERN, European Organization for Nuclear Research, Geneva, Switzerland\\
% ...

\end{sloppypar}
\end{document}